\documentclass[preprint]{elsarticle}

\usepackage{euscript,amssymb,amsmath}

\usepackage{graphicx}
\usepackage{epsfig}
\usepackage{color}

\newcommand{\be}[1]{\begin{equation}\label{#1}}
\newcommand{\ee}{\end{equation}}
\newcommand{\ba}[1]{\begin{eqnarray}\label{#1}}
\newcommand{\ea}{\end{eqnarray}}
\newcommand{\rf}[1]{(\ref{#1})}
\newcommand{\nn}{\nonumber}

\newcommand{\de}{\partial}

\newcommand{\la}{\lambda}

\newcommand{\bt}{\beta}

\newcommand{\hg}{\hat g}
\newcommand{\hR}{\hat R}
\newcommand{\hT}{\hat T}
\newcommand{\ka}{\kappa}
\newcommand{\La}{\Lambda}
\newcommand{\mcd}{\mathcal{D}}
\newcommand{\ve}{\varepsilon}
\newcommand{\om}{\omega}
\newcommand{\Om}{\Omega}
\newcommand{\td}{\tilde}
\newcommand{\hna}{\hat\nabla}
\newcommand{\tr}{\triangle}
\newcommand{\mf}{\mathfrak}

\newcommand{\re}{{\bf r}}

\newcommand{\Ue}{U_{\rm eff}}
\newcommand{\im}{\imath}

\begin{document}

\begin{frontmatter}

\title{Problematic aspects of Kaluza-Klein excitations in multidimensional models with Einstein internal spaces}

\author[a,b]{Alexey Chopovsky}
\ead{a.chopovsky@yandex.ru}

\author[c]{Maxim Eingorn}
\ead{maxim.eingorn@gmail.com}

\author[b]{Alexander Zhuk}
\ead{ai.zhuk2@gmail.com}

\address[a]{Department of Theoretical Physics, Odessa National University,\\ Dvoryanskaya st. 2, Odessa 65082, Ukraine}

\address[b]{Astronomical Observatory, Odessa National University,\\ Dvoryanskaya st. 2, Odessa 65082, Ukraine}

\address[c]{CREST and NASA Research Centers, North Carolina Central University,\\ Fayetteville st. 1801, Durham, North Carolina 27707, U.S.A.\\}

%
%
%

%
\begin{abstract} We consider Kaluza-Klein (KK) models where internal spaces are compact Einstein spaces. These spaces are stabilized by background matter
(e.g., monopole form-fields). We perturb this background by a compact matter source (e.g., the system of gravitating masses) with the zero pressure in the
external/our space and an arbitrary pressure in the internal space. We show that the Einstein equations are compatible only if the matter source is smeared
over the internal space and perturbed metric components do not depend on coordinates of extra dimensions. The latter means the absence of KK modes
corresponding to the metric fluctuations. Maybe, the absence of KK particles in LHC experiments is explained by such mechanism.
\end{abstract}
%



\begin{keyword} multidimensional models \sep Kaluza-Klein models \sep Kaluza-Klein excitations/modes \sep Einstein spaces \sep black strings/branes \sep
gravitational tests \end{keyword}

\end{frontmatter}



\section{Introduction}

\setcounter{equation}{0}

One of the main tasks of the  Large Hadron Collider (LHC)  is to search for physical phenomena beyond the standard model. If such events occur, they should
suggest the directions of the further development of physics. Extra spatial dimensions are among these phenomena. The idea of multidimensionality of our
Universe demanded by the theories of unification of the fundamental interactions is one of the most breathtaking ideas of theoretical physics. It takes its
origin from the pioneering papers by Th. Kaluza and O. Klein \cite{KK}, and now the most self-consistent modern theories of unification such as superstrings,
supergravity and M-theory are constructed in spacetimes with extra dimensions (see, e.g., \cite{Polchinski}). One of the clearest evidence for the existence of
extra dimensions is the detection of Kaluza-Klein (KK) modes/particles which correspond to appropriate eigenfunctions of the internal manifold, i.e.
excitations of fields in the given internal space. Such excitations were investigated in many articles (see, e.g., the classical papers \cite{SS,Nieu,Kim}).
However, up to now, such particles were not detected in the LHC experiments (see, e.g., \cite{ATLAS,ATandCMS}). In our letter we present a possible explanation
of this problem in the case of KK models where stabilized internal spaces are compact Einstein spaces. We perturb this multidimensional background by
gravitating masses and show that multidimensional Einstein equations are consistent only if the gravitating bodies are uniformly smeared over the internal
space and, consequently, the metric fluctuation components depend only on the coordinates of the external/our space. Obviously, this means the absence of KK
modes corresponding to the metric fluctuations. We also demonstrate that the coupling constant between gravitating masses and radions depends on the equation
of state parameter $\Omega$ for the gravitating masses in the internal space. In the case of black strings/branes ($\Omega=-1/2$) the coupling disappears and
the gravitating masses do not perturb the internal space. This explains the excellent agreement of black strings/branes with the gravitational tests.

\


\section{Smearing over extra dimensions}

Let
\be{1}
\hg_{MN}(y)dX^M\otimes dX^N=\eta_{\mu\nu}dx^{\mu}\otimes dx^\nu+\hg^{(d)}_{mn}(y)dy^m\otimes dy^n
\ee
be a metrics on $(\mathcal{D}=1+D=4+d)$-dimensional background spacetime manifold ${\mf M}_{\mcd}={\mf M}_4\times{\mf M}_{d}$, where ${\mf M}_4$ is the
Minkowski spacetime and  the internal space ${\mf M}_{d}$ is a compact Einstein one:
\be{2}
\hR_{mn}[\hg^{(d)}]=\la\hg_{mn},\quad\hR^m_m[\hg^{(d)}]=\hR^{(d)}=\la d, \quad
\la\equiv{\rm const}. \ee
In general, ${\mf M}_{d}$ can be an orbifold. In what follows, $M=0,1,2,\ldots ,D$; $\mu = 0,1,2,3$; $\tilde \mu = 1,2,3$ and $m =4,5,\ldots ,3+d$. According
to our sign convention, $\lambda<0$ ($\lambda >0$) corresponds to the compact Einstein space with positive (negative) curvature. For example, in a particular
case of the $d$-dimensional sphere of the radius $a$ we have $\lambda = -(d-1)/a^2$. The case of a Ricci-flat internal space $\lambda=0$ and its particular
example of the toroidal compactification were studied in our previous papers \cite{ChEZ5,ChEZ6} and will not be considered here.

Obviously, to create the curved background spacetime with the metrics \rf{1}, we should introduce also the background matter. The properties of the
energy-momentum tensor (EMT) of this matter can be easily determined from the Einstein equations \cite{notation}:
\be{3}
\kappa \hT '^M_N=\hR^M_N-\left(\frac{1}{2}\,\hR^{(d)} +\kappa \La_{\mcd}\right)\delta^M_N, \quad \ka\equiv\frac{2S_D\td  G_{\mcd}}{c^4}\,,
\ee
where $\La_{\mcd}$ is the multidimensional cosmological constant, $S_D=2\pi^{D/2}/\Gamma (D/2)$ is the total solid angle, $\tilde G_{\mathcal{D}}$ is the
gravitational constant in the $(\mathcal{D}=D+1)$-dimensional spacetime. For the metrics \rf{1} we get
\be{4}
\hT '^\mu_\nu=-\left( \frac{\la d}{2\ka}+\La_\mcd \right)\delta^\mu_\nu, \quad \hT '^m_n=-\left( \frac{\la (d-2)}{2\ka}+\La_\mcd \right)\delta^m_n.
\ee
It is worth noting that these equations coincide with Eqs. (11) and (12) in \cite{CandWein} in the case of Einstein internal spaces. It was indicated there
that the multidimensional cosmological constant $\La_\mcd$ is necessary to get a solution with the flat external space. In section 3, we will show additionally
that we need this constant to ensure the stable compactification of the internal space. We can rewrite these expressions in the form of the components of a
perfect fluid:
\be{5}
\hT '^M_N={\rm diag}\bigl(\,\hat \ve', -\hat p'_0, -\hat p'_0, -\hat p'_0, \underbrace{-\hat p'_1, ... , -\hat p'_1}_{d\,\, {\rm times}}\,\,\bigr),
\ee
\be{6} \hat \ve'\equiv -\left[ (\la d/2\ka)+\La_\mcd \right], \quad \hat p'_0=\om_0 \hat \ve', \quad \hat p'_1=\om_1 \hat \ve'\, , \ee
where the parameters of equations of state are
\be{7}
\om_0=-1, \quad \om_1=[(2-d)\la-2\ka\La_\mcd]/(\la d+2\ka\La_\mcd)\,.
\ee
Here, $\omega_0=-1$ corresponds to the vacuum-like equation of state in the external space, and in the internal space the parameter $\omega_1$ is an arbitrary.
Choosing different values of $\omega_1$ (with fixed $\omega_0 = -1$), we can simulate different forms of matter. For example, $\omega_1=1$ corresponds to the
monopole form-fields (the Freund-Rubin scheme of compactification \cite{FR,AGHK,GMZ}\footnote{Different compactification schemes (i.e. choices of the form of
the electromagnetic field) may result in different values of $\omega_1$ \cite{GRT}.}).  For the Casimir effect we have $\omega_1=4/d$ \cite{CandWein,exci}.
Obviously, for these forms of matter we can calculate the energy density and pressures in the external and internal spaces, and Eqs. \rf{6} and \rf{7} should
be treated as the fine tuning conditions.

Eqs. \rf{6} and
\rf{7} result also in the following useful auxiliary relation:
\be{8}
\hat \ve'=-\la/[\ka(1+\om_1)]\, .
\ee

Now, we perturb the above background by a perfect fluid with the following EMT (see also \cite{ChEZ6}):
\begin{eqnarray}\label{9}
  \td  T^{M \nu}&=&\tilde \rho^{(\mcd)} c^2 \frac{ds}
{dx^0} u^M u^\nu,\quad u^M=\frac{dX^M}{ds},\nn\\
  \td  T^{mn}&=&-\td  pg^{mn}+\tilde \rho c^2 \frac{ds}
{dx^0} u^m u^n, \quad \td  p=\Om\tilde \rho^{(\mcd)} c^2\frac{ds}{dx^0}\, .
\end{eqnarray}
This perfect fluid is pressureless in the external/our space but it has pressure in the internal space. $\Omega$ is the equation of state parameter in the
internal space. Usually we assume that this matter source is compact. For example, such EMT can corresponds to a system of gravitating bodies, e.g., ordinary
astrophysical objects. It is well known that the pressure inside of these masses (e.g., inside of our Sun) is much less than the energy density. Therefore, we
can neglect it. However, we do not know the pressure of these bodies in the internal space. So, we keep it. We should note that this pressure is not connected
with motion of gravitating masses, i.e. $\Omega$ is the parameter of a body. Obviously, in the case of a gravitating body with the mass $m$ the rest mass
density is
\be{10} \tilde \rho^{(\mcd)} = [|g|]^{-1/2} m \delta (\bf{X})\, . \ee
We can easily generalize this expression to the case of a system of gravitating masses \cite{ChEZ6}.  For our purpose, it is sufficient to consider a
steady-state model. For example, we can consider only one gravitating mass and place the origin of a reference frame on it. That is, we disregard the spatial
velocities of the gravitating masses.

It is worth noting that the metric components $g_{MN}$ in Eqs. \rf{9} and \rf{10} are perturbed ones, and in the weak-field approximation up to $O(1/c^2)$
correction terms they can be written in the form
\be{11}
g_{MN}\approx \hg_{MN}+h_{MN} \, ,
\ee
where correction terms $h_{MN}\sim O(1/c^2)$ can be found with the help of the Einstein equation:
\be{12}
\frac{1}{\ka}\,R_{MN}=T_{MN}-\frac{1}{d+2}\,Tg_{MN}-\frac{2}{d+2}\,\La_{\mcd}g_{MN}\, .
\ee
The total EMT in the right hand side of this equation is $T_{MN}=T '_{MN}+\td  T_{MN}$, where $T '_{MN}$ is the EMT of the perturbed background matter. To get
the metric correction terms $h_{MN}\sim O(1/c^2)$, we should determine the components of $T_{MN}$ up to $O(c^2)$ terms. For example, the components \rf{9} are
approximated as
\ba{13}
\td  T_{00}&\approx& \rho^{(\mcd)} c^2,\quad \td  T_{\td \mu\td \nu}\approx0,\quad \td  T_{0 M}\approx0,\nn \\
\td  T_{mn}&\approx&-\Om \rho^{(\mcd)} c^2
\hg_{mn},\quad \td  T\approx \rho^{(\mcd)} c^2(1-\Om d),
\ea
where $\rho^{(\mcd)}$ denotes the rest mass density (see \rf{10}) with respect to the unperturbed metrics $\hat g_{MN}$.

Concerning the EMT of the background matter, we suppose that perturbation does not change the equations of state in the external and internal spaces, i.e.
$\omega_0$ and $\omega_1$ are constants. For example, if we had monopole form-fields ($\omega_0=-1,\,\omega_1=1$) before the perturbation, the same type of
matter we have after the perturbation. Thus, the EMT of the perturbed background is
\be{14} T '_{\mu\nu}\approx \left(\hat \ve '+\ve'_1\right)g_{\mu\nu}\, , \quad T '_{mn}\approx -\omega_1\left(\hat \ve '+\ve'_1\right)g_{mn}\, ,\ee
where the correction $\ve'_1$ is of the same order of magnitude as the perturbation $\rho^{(\mcd)} c^2$.

According to the EMT \rf{9}, the pressure is isotropic in each factor manifolds. Obviously, such perturbation does not change the topology of the model, i.e.
the topologies of the factor manifolds. Additionally, it preserves also the block-diagonal structure of the metric tensor. In the case of a steady-state model
(our case) the non-diagonal perturbations $h_{0{\tilde M}}$ are also absent. Therefore, the metric correction terms are conformal to the background metrics and
can be written in the block-diagonal form:
\be{15} \left[h_{MN}(X)\right]=\left[\xi_1 \eta_{00}\right]\otimes\left[\xi_2 \eta_{\td \mu\td \nu}\right]\otimes\left[ \xi_3 \hg^{(d)}_{mn}\right]\,, \ee
where $\xi_{1, 2, 3}=\xi_{1, 2, 3}(X)\sim O(1/c^2)$. Then, the EMT \rf{14} is approximated up to $O(c^2)$ correction terms as follows:
\ba{16}
T'_{00}&\approx&\hat \ve'(1+\xi_1)+\ve'_1\, ,\quad
T'_{\td \mu\td \nu}\approx\left[\hat \ve'(1+\xi_2)+\ve'_1\right]\eta_{\td \mu\td \nu}\, ,\nn\\
T'_{mn}&\approx&-\om_1\left[\hat \ve'(1+\xi_3)+\ve'_1\right]\hg^{(d)}_{mn}\, ,\\
T'&\approx&(4-\om_1 d)(\hat \ve'+\ve'_1)\, .\nn
\ea

Taking into account the reasoning in Appendix A \cite{ChEZ6} as well as the background relations \rf{7} and \rf{8} above, the Einstein equation \rf{12} reads
(up to $O(1/c^2$) terms):
\ba{17}
\tr_D\xi_1&=&2\frac{1+d(\Om+1)}{d+2}\ka\tilde\varepsilon+2\frac{d(1+\om_1)-2}{d+2}\ka\ve'_1, \\
\tr_D\xi_2&=&-2\frac{1-\Om d}{d+2}\,\ka\tilde\varepsilon+2\frac{d(1+\om_1)-2}{d+2}\,\ka\ve'_1,\label{18}\\
\tr_D\xi_3&=&-2\frac{1+2\Om}{d+2}\,\ka\tilde\varepsilon+2\la\xi_3-\frac{4(\om_1+2)}{d+2}\,\ka\ve_1'\, ,\label{19}
\ea
where $\tilde\varepsilon \equiv \rho^{(\mcd)}c^2$ and $\tr_D$ is the $D$-dimensional Laplace operator with respect to the metrics $\hg_{MN}$. To get this
system, we used the well known  gauge condition \cite{Landau}
\be{20}
\hna_Lh^L_N-\frac{1}{2}\,\de_Nh^L_L=0,\quad h^M_N\equiv\hg^{MS}h_{NS}\, ,
\ee
which in our case is reduced to the system
\ba{21}
\de_0\xi_1&-&(1/2)\,\de_0\left(\xi_1+3\xi_2+\xi_3 d\right)=0\, ,\\
\de_{\td\nu}\xi_2&-&(1/2)\,\de_{\td\nu}\left(\xi_1+3\xi_2+\xi_3 d\right)=0\, ,\label{22}\\
\de_{n}\xi_3&-&(1/2)\,\de_{n}\left(\xi_1+3\xi_2+\xi_3 d\right)=0\label{23}\, . \ea
This condition simplifies considerably the Einstein equations. Obviously, it does not affect the physical results. The condition \rf{21} is satisfied
automatically for the stationary perturbations $\xi_i$. Also, for the stationary perturbations $\xi_i$ and the compact matter source with the EMT \rf{9}, the
conditions \rf{22} result in the following:
\be{24} \xi_1+\xi_2+\xi_3d=\biggl.C(y)\biggr|_{|\re| \rightarrow+\infty}\rightarrow0\quad\Rightarrow\quad \xi_3=-\frac{1}{d}\left(\xi_1+\xi_2\right), \ee
where we took into account that at large distances $|\re|$ from the compact matter source the metrics goes asymptotically to the background one. Hereafter,
$\re$ denotes the radius-vector in the external three-dimensional space.

Now, applying the Laplace operator $\tr_D$ to the relation \rf{24} and comparing it with Eqs. \rf{17}-\rf{19}, first, we get
\be{25}
\ve'_1=(\la d/2\ka)\,\xi_3
\ee
and, second, Eq. \rf{19} takes the form
\be{26} \tr_D \xi_3=2\left[ \la\,\frac{2-d(1+\om_1)}{d+2}\,\xi_3-\frac{1+2\Om}{d+2}\,\ka\rho^{(\mcd)}c^2 \right]. \ee
The relation \rf{24} shows that functions $\xi_\im (\im=1,2,3)$ are not linearly independent. So, $\xi_{1,2}$ can be presented as
\be{27}
\xi_\im=-(d/2) \xi_3+\beta_\im f, \quad \im=1,2\, ,
\ee
where $f$ is a new function and constant parameters $\beta_i$ satisfy the condition $\beta_1+\beta_2=0$. Then, taking into account Eq. \rf{26}, we can easily
show that Eqs. \rf{17} and \rf{18} are reduced now to the system
\be{28}
\beta_1\tr_D f=\,\ka\rho^{(\mcd)}c^2 , \quad\beta_2\tr_D f=-\,\ka\rho^{(\mcd)}c^2\, .
\ee
Therefore, we can put
\be{29}
\beta_1=-\beta_2=1 \quad \Rightarrow \quad \tr_D f=\,\ka\rho^{(\mcd)}c^2
\ee
and, hence, for the functions $\xi_1$ and $\xi_2$  we get
\be{30}
\xi_1=-(d/2)\,\xi_3+f, \quad \xi_2=-(d/2)\,\xi_3-f\, .
\ee
Now, coming back to the gauge condition \rf{23}, we get
\be{31}
(d+2)\xi_3+2f=C_1(\bf r)\, ,
\ee
where $C_1(\bf r)$ is a function that depends only on the external coordinates. Applying to this relation the Laplace operator $\tr_D$ and using Eqs. \rf{26}
and \rf{29}, we arrive at the following crucial equation:
\be{32}
\la\left[2-d(1+\om_1)\right]\xi_3-2\Om\ka\rho^{(\mcd)} c^2=C_2({\bf r})/2\, ,
\ee
where $C_2({\bf r}) = \tr_D C_1(\bf r)$. As we will see below, the condition of stable compactification leads to the requirement
$\la\left[2-d(1+\om_1)\right]>0$. Then, from \rf{32} we conclude, that in the region outside the compact matter source with the EMT \rf{9} (i.e. where
$\rho^{(\mcd)}=0$), $\xi_3$ depends only on the external coordinates. Hence, the function $\rho^{(\mcd)}$, and, therefore, also the function $f$ depend only on
the external coordinates. This means that the matter source with the EMT \rf{9} (e.g., a system of gravitating masses) should be uniformly smeared over the
internal space. In this case the $D$-dimensional Laplacian $\tr_D$ in Eqs. \rf{26} and \rf{29} should be replaced by a three-dimensional one $\tr_3$ (with
respect to the flat metrics), and the multidimensional rest mass density $\rho^{(\mcd)}$ is reduced to a three-dimensional one: $\rho^{(\mcd)}=\rho^{(3)}({\bf
r})/V_{\mathrm{int}}$ where $V_{\mathrm{int}}$ is the unperturbed internal space volume. Thus, Eqs. \rf{29} and \rf{26} read
\ba{33}
\tr_3 f&=&\frac{8\pi G_N}{c^2}\rho^{(3)}\, ,\\
\tr_D \xi_3-\mu^2\, \xi_3 &=& - \frac{2(1+2\Omega)}{d+2}\frac{8\pi G_N}{c^2}\rho^{(3)}\, ,\label{34}
\ea
where we introduced the Newton gravitational constant $4\pi G_N = S_D\td  G_{\mcd}/V_{\mathrm{int}}$ and the Yukawa radion mass squared $\mu^2 \equiv
2\lambda[2-d(1+\om_1)]/(d+2)$. In our case, the Yukawa scalar particle is a radion/gravexciton \cite{exci}. It is well known that to get the physically
reasonable solution of \rf{34} with the boundary condition $\xi_3 \to 0$ for $|\bf r| \to +\infty$ the parameter $\mu^2$ should be positive. This results in
inequalities:
\be{35}
\la\left[2-d(1+\om_1)\right]>0 \; \Rightarrow\; \left\{
   \begin{array}{cc}
   \omega_1 >(2/d) -1, & \lambda <0,\\
   \omega_1 <(2/d) -1, & \lambda >0.
   \end{array}
\right.
\ee
In the case of a point-like source $\rho^{(3)}({\bf r})=m \delta({\bf r})$, and the solutions of Eqs. \rf{33} and \rf{34} are
\be{36}
f=\frac{2\varphi_N}{c^2},\quad \xi_3 = - \frac{4\varphi_N}{(2+d)c^2}(1+2\Omega)\exp(-\mu |{\bf r}|)\,,
\ee
where the Newtonian potential $\varphi_N=-G_Nm/|{\bf r}|$. If the internal space is a $d$-dimensional sphere of the radius $a$ (i.e. $\lambda = -(d-1)/a^2$),
then the solutions \rf{36} exactly coincide with the ones in \cite{EFZ2}. We can conclude from \rf{36} that the Yukawa coupling constant $g$ between any
massive particle and the radion is determined as $g^2 \sim (d/(2+d))G_N (1+2\Omega)$. In contrast to the radion mass $\mu \sim 1/a$, the coupling constant does
not depend on the size of the internal space. Moreover, in the case of black branes $\Omega =-1/2$ \cite{ChEZ3}, the coupling disappears at all. As a result,
the presence of a gravitating mass does not excite the internal space: $\xi_3 \equiv 0$ (see Eq. \rf{36}).


\

\section{Internal space stabilization}

Let us show now that inequalities \rf{35} correspond to the internal space stability conditions. To prove it, we suppose, that the scale factor of the internal
space is a function of time:
\be{37}
\hg^{(d)}_{mn}(y)\quad \longrightarrow\quad e^{2\bt(t)}\hg^{(d)}_{mn}(y), \quad t\equiv x^0\, .
\ee
Without loss of generality, we may put $\beta(t=t_0)=0$ where $t_0$ is the present time. Then, from the conservation law $T'^M_{N;\,M}=0$ for the diagonal EMT
\rf{5} ($\omega_0=-1$), we get
\be{38}
\ve'(t)=\ve_c' e^{-\bt(1+\om_1)d}\, ,
\ee
where $\ve_c'$ is a constant of integration. The stabilization of the internal space is possible if the effective potential (see for details \cite{exci,Zhuk})
\be{39}
\Ue(\bt)=e^{-d\bt}\left[ (\hR^{(d)}/2)e^{-2\bt}+\ka\La_{\mcd}+\ka\ve'_ce^{-\bt(1+\om_1)d} \right]
\ee
has a minimum at $t=0$. Since in the considered model the external space is flat, $\La^{(4)}_{\rm eff}=\Ue(\bt=0)=0$, then
\be{40} \hR^{(d)}=\la d=-2\ka\left(\La_{\mcd}+\ve'_c\right)\, , \ee
that coincides with \rf{6} for $\ve'_c=\hat\ve'$. The necessary condition for an extremum of the effective potential
$\left.\de\Ue/\de\bt\right|_{\bt=0}=0$ leads to the relation
\be{41} \la=-(1+\om_1)\ka\hat\ve'\, , \ee which exactly coincides with \rf{8}. The sufficient condition of the minimum $\left.\de^2\Ue/\de\bt^2\right|_{\bt=0}
>0$ results in inequality
\be{42}
\la \left[2 - d(1+\om_1)\right] >0\, .
\ee
Obviously, this inequality coincides with condition \rf{35} of the positiveness  of the Yukawa radion mass squared. If we also demand the positivity of the
unperturbed background energy density $\hat \ve' $ determined in \rf{8}, then we finally get $\omega_1 >(2/d)-1$ for $\lambda <0$ and  $\omega_1 <-1$ for
$\lambda >0$. Additionally, it can be easily seen that in the case $\La_\mcd=0$ we have $\omega_1 = (2-d)/d \; \Rightarrow\;
\left.\de^2\Ue/\de\bt^2\right|_{\bt=0} = \mu^2 =0$. Therefore, the presence of $\La_\mcd$ is the necessary condition for the internal space stabilization.


\

\section{Summary}

In this letter, we have considered KK models where internal spaces are compact Einstein spaces, e.g., orbifolds. These spaces are stabilized by background
matter, e.g., monopole form-fields. We perturbed this background by a compact matter source with the zero pressure in the external/our space and an arbitrary
pressure in the extra dimensions. For example, such matter source can be modeled by a gravitating mass. Then, we investigated the metric perturbations in the
weak-field limit and showed that the Einstein equations are compatible only if both the metric perturbations and the energy density of the compact matter
source do not depend on the coordinates of the internal space. For gravitating masses, this means their smearing over the internal space. For the metric
perturbations, this means the absence of the KK modes. The KK modes/particles are the object of the active search in LHC experiments. However, up to now such
modes are not detected \cite{ATLAS,ATandCMS}. As we show in our letter, the reason for this may be smearing of particles and fields over the internal spaces,
although it looks unnatural from the point of view of statistical physics because any nonzero temperature should result in fluctuations, i.e. in KK states.

It should be mentioned that the metric and field perturbations are often considered without taking into account the reason of such fluctuations (see, e.g.,
\cite{KLZ,BD}). Our analysis clearly shows that the inclusion of the matter sources, being responsible for the perturbations, imposes strong restrictions on
the model. Moreover, these restrictions are not caused directly by the made choice of the energy-momentum tensors of the sources. For instance, KK excitations
do exist under a similar choice with $\Omega=0$ in the case of toroidal compactification \cite{EZ1}, but then there is a contradiction with the relativistic
gravitational tests \cite{EZ3}.

Besides, it is worth noting that our investigation is related also to the popular Universal Extra Dimension models \cite{theor bounds UED} if the internal
spaces are orbifolds. Clearly, the brane world models require a separate investigation. There are also models with combined KK and brane topology \cite{RS+KK}.
Our research shows that in such combined models the smearing of gravitational sources can also occur.

\


\section*{Acknowledgements}
The work of M. Eingorn was supported by NSF CREST award HRD-1345219  and NASA grant NNX09AV07A. A. Zhuk acknowledges the hospitality of the Theory Division of
CERN during the final preparation of this paper.




\begin{thebibliography}{}
\bibitem{notation}
According to our notations, the hat indicates unperturbed quantities (metric coefficients and matter) and the apostrophe denotes the quantities that
characterize the background matter (both unperturbed and perturbed).
\end{thebibliography}
\end{document}